\documentstyle[fleqn,psfig]{tp}

\def\ap3m{\mbox{AP$^3$M}}

\def\lcdm30art{\mbox{\char'3CDM$_{30}^{\rm ART}$~}}

\def\sig8{\mbox{$\sigma_8$}}

\def\lesssim{\mbox{$_ <\atop{^\sim}$}}
\def\gtrsim{\mbox{$_ >\atop{^\sim}$}}

\begin{document}

\twocolumn[
\title{Evolution of isolated halos and halos inside of groups and
clusters in a $\Lambda$CDM model}
\author{S. Gottl\"ober$^{1,2}$, A.A. Klypin$^2$, A.V. Kravtsov$^2$\\[3mm]}
{\it $^1$AIP, An der Sternwarte 16, D-14482 Potsdam}\\ {\it
$^2$Astronomy Department, NMSU, Dept.4500, Las Cruces, NM 88003-0001}\\
\vspace*{16pt}   

ABSTRACT.\ A significant fraction of mass in the universe is believed
to be in the form of dark matter (DM). Due to gravitational
instability, the DM collapses hierarchically into DM {\em halos}. In
this contribution we present a study of the formation and evolution of
such DM halos in a {\sl COBE}-normalized spatially flat $\Lambda$CDM
model ($\Omega_0=1-\Omega_{\Lambda}=0.3$; $h=0.7$) using
high-resolution $N$-body simulations. The novelty of this study is use
of the newly developed halo-finding algorithms to study the evolution
of both {\em isolated} and {\em satellite} (located inside virial radii
of larger group- and cluster-size systems) halos. The force and mass
resolution required for a simulated halo to survive in the high-density
environments typical of groups and clusters is high: $\sim 1-3$ kpc and
$\sim 10^9{\rm M_{\odot}}$, respectively. We use the high-resolution
Adaptive Refinement Tree (ART) $N$-body code to follow the evolution of
$256^3$ dark matter particles with dynamic range in spatial resolution
of $32,000$ in a box of $60 h^{-1}$ Mpc.

We show that the correlation function of these halos is anti-biased with
respect to the dark matter correlation function and is high and 
steeper than the correlation function of the isolated 
virialized objects. The correlation function evolves only mildly
between $z=3$ and $z=1$. The mass evolution of isolated virialized
objects determined from the simulation is in good agreement with 
prediction of semi-analytical models. The differences exist, however,
if we include satellite halos in the halo catalogs.

\endabstract]

\markboth{S. Gottl\"ober et al.}{Evolution of halos}

\small

\section{Introduction}

It is generally believed that dark matter (DM) constitutes a large
fraction of the mass in the Universe. Therefore, it significantly
affects both the process of galaxy formation and the large-scale
distribution of galaxies. The most convincing observational evidence
for substantial amounts of dark matter even in the very inner regions
of galaxies comes from recent HI studies of dwarf and low surface
brightness galaxies. The gravitational domination of DM on the scale of 
galaxy virial radius implies that collisionless simulations can be used to
study formation of the DM component of galaxies. 

A well known problem of dissipationless simulations is overmerging,
i.e. the lack of substructure in virialized objects that could be
associated with galaxy locations. This effect is due mainly to
insufficient force and mass resolution (Moore et al. 1996; Klypin et
al. 1998, hereafter KGKK). Recently, the dynamic range of the $N$-body
simulations has become sufficiently high to overcome this problem. 
With sufficient resolution the DM halos
survive and can be identified in high-density regions even in
collisionless simulations (KGKK; Ghigna et al. 1998). This allows us to
study halo dynamics and the physical effects of tidal stripping and
dynamical friction in group- and cluster-size objects.

\begin{figure*}
 \centerline{\psfig{figure=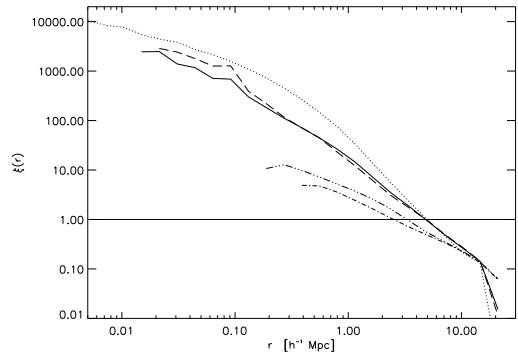,height=5cm}}
\caption[]{The correlation function of halos with circular velocities
$v_{circ} > 120$ km/s (solid line) and $v_{circ} > 150$ km/s 
(dashed line) in comparison to the correlation function of dark
matter particles (dotted line). The dashed-dotted lines represent the
correlation function of objects found by the friends-of-friends
algorithm with the linking radius corresponding to the virial overdensity
(dash-dotted for objects with $N > 100$ particles and dash-triple-dotted for
 $N > 20$).}
\label{fig:1}
\end{figure*}

In this contribution, we discuss the evolution of DM halos in a
spatially flat cosmological model dominated by cold dark matter and
non-zero cosmological constant ($\Lambda$CDM). Specifically, we study
the differences between isolated and satellite halos in their mass
evolution and spatial distribution. Scenarios with a $\Lambda$-term
have become very successful in describing most aspects of cosmological
structure formation.  We have chosen the following parameters in order
to reconcile the model with both the {\sl COBE} measurements and cluster
abundance: $\Omega_0=1-\Omega_{\Lambda}=0.3$; $h=0.7$; $\sigma_8=1.0$.
The age of the universe in this model is $\approx 13.5$ Gyrs.

\section{Numerical Simulation}

In order to study the properties of halos in a cosmological
environment, the simulation box should be sufficiently large. On the
other hand, to assure that halos do survive also in clusters, the force
resolution should be $\sim 1-3h^{-1} {\rm kpc}$ and the mass resolution
should be $\lesssim 10^9h^{-1} {\rm M_{\odot}}$ (Moore et al. 1996;
KGKK).  The Adaptive Refinement Tree (ART) $N$-body code (Kravtsov,
Klypin \& Khokhlov 1997) reaches a formal dynamical range of $32,000$
in high density regions, which for the 60 $h^{-1}$ Mpc box corresponds
to the required necessary force resolution. In the 60 $h^{-1}$ Mpc box
with $256^3$ particles, each particles has a mass of $1.1 \times 10^9
h^{-1} {\rm M_{\odot}}$.

Identification of halos in dense environments and reconstruction of
their evolution is a challenge. Most widely used halo-finding
algorithms, the friends-of-friends (FOF) and the spherical overdensity,
both discard ``halos inside halos'', i.e. satellite halos located
within virial radius of larger halos. The distribution of halos
identified in this way, cannot be compared easily to the distribution
of galaxies, because the latter are found within larger systems. In
order to cure this, we have developed two related algorithms, which we
have called {\em hierarchical friends-of-friends} (HFOF) and {\em bound
  density maxima} (BDM) algorithms (KGKK). These algorithms are
complementary.  Both find essentially the same halos. The
advantage of the HFOF algorithm is that it can handle halos of
arbitrary (not only spherically symmetric) shape. The advantage of the
BDM algorithm is that it separates background unbound particles
from the particles gravitationally bound to the halo,  and 
thus allows a better determination of physical properties of halos.

Since the algorithms work on a snapshot of the particle distribution,
they tend to identify also small fake ``halos'' consisting of only a
few {\em unbound} particles, clumped together by chance at the analyzed
moment. We deal with this problem by both checking whether the
identified clump is gravitationally bound and by following the merging
history of halos. Halos that do not have a progenitor at a previous
moment are discarded. For other halos we find the direct progenitor,
i.e.  a halo at a previous moment that contains the maximum number of
particles of this halo. We use the chain of progenitors identified in
this way to reconstruct the mass evolution of a given halo back in
time, down to the epoch of its first detection in the simulation.

About 10,000 halos (with maximum circular velocity $\gtrsim 120$ km/s, cf.
Kravtsov et al. in this volume) can be identified at $z=0$ in the
analyzed 60 $h^{-1}$ Mpc simulation box.  Several hundreds of these
halos are located in groups and clusters.  This allows us to carry out
for the first time a statistical comparison study of the clustering and
mass evolution of isolated and satellite galaxy-size halos in an
hierarchical cosmological model.

\section{Results}

We have identified halos using different input numerical parameters
(size and number of particles) for the halo-finding algorithms. We find
that for distances $\gtrsim 100 h^{-1}$ kpc the resulting halo-halo
2-point correlation function (CF) does not depend on these assumptions,
as long as the number of particles in identified halo is $\gtrsim 30$.
Note that the halo radius does not set limits on the inter-halo
separation because halos are allowed to overlap. In Fig. 1 we present
the halo-halo correlation function for halos with a maximum radius of
$100 h^{-1}$ kpc and more than 28 particles ($3\times 10^{10} h^{-1}
{\rm M_{\odot}}$). There is a statistically significant anti-bias on
scales less than the correlation radius $r_0$. The slope of the CF is
$\approx -1.8$ at scales $\approx 1-15 h^{-1} {\rm Mpc}$; at smaller
scales the correlation function flattens slightly. In agreement with
Col\'{\i}n et al. (1998) and results of many other studies, we find
almost no time evolution of the CF at $z<3$. The correlation function
of DM, on the other hand, evolves rapidly which results in the strong
evolution of bias and transition from $b>1$ to $b<1$ at small scales at
$z\approx 1$ (see Fig. 2, see also Kravtsov et al. on the bias
evolution in this volume).

The correlation function for the halos identified using our
algorithms is compared in Fig.1 to the CF of halos identified in a
standard way using the FOF algorithm with linking radius corresponding
to the virial overdensity. As we noted above, by definition the
satellite halos are not present in the catalogs so generated. The
comparison shows clearly that there are significant differences in both
the slope and the amplitude at small ($\lesssim 7h^{-1} {\rm Mpc}$)
scales between these correlation functions. The higher amplitude of the
CF determined using our halo-finding algorithms is explained by the
larger number of small-separation halo pairs formed by satellite halos
in groups and clusters missing in the FOF catalog. Note also that the
CF of FOF halos does not extend down to $200h^{-1} {\rm kpc}$ due to
the self-exclusion of halos at separations smaller than halo virial
radii.

\begin{figure}
 \centerline{\psfig{figure=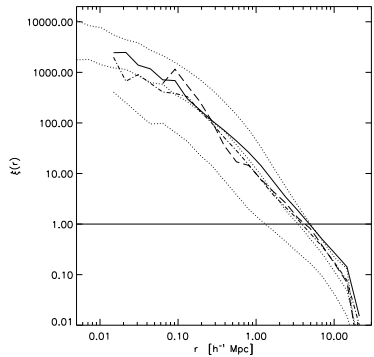,height=5cm}}
\caption[]{The correlation function of halos with circular velocities
$v_{circ} > 120$ km/s at $z = 0$ (solid line), $z = 1$ (dash-dotted
line), and $z=3$ (dashed line) in comparison to the correlation
function of dark matter particles at the same redshifts (dotted lines;
amplitude increases with decreasing redshift).
}
\label{fig:2}
\end{figure}

The mass of an object found by the HFOF algorithm {\em at virial overdensity}
can be defined as the sum of linked particle masses.  For all of the HFOF
objects we identify the main progenitors at all epochs down to the halo
formation time. To study the mass evolution due to merging and
accretion we have divided these objects into four mass bins at $z=0$: $
M_0 > 5 \times 10^{12}{\rm M_{\odot}}$ (bin 1), $5 \times 10^{12}{\rm
  M_{\odot}} > M_0 > 5 \times 10^{11}{\rm M_{\odot}}$ (bin 2), $ 5
\times 10^{11}{\rm M_{\odot}} > M_0 > 5 \times 10^{10}{\rm M_{\odot}}$
(bin 3), and $ M_0 < 5 \times 10^{10}{\rm M_{\odot}}$ (bin 4). Average
mass evolution of halos in these bins normalized to the mass at $z = 0$
is shown in Fig. 3. We find a good agreement with the semi-analytical
predictions (Lacey \& Cole 1993) for the evolution of these {\em isolated} halos.

\begin{figure}
 \centerline{\psfig{figure=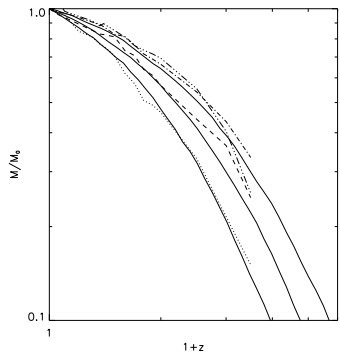,height=5cm}}
\caption[]{Mass evolution of halos identified by the HFOF algorithm
  at virial overdensity. The dotted line is for average halo mass of
  $1.4 \times 10^{13}{\rm M_{\odot}}$ (10 halos), the dashed for $1.1
  \times 10^{12}{\rm M_{\odot}}$ (20 halos), the dash-dotted for $7.7
  \times 10^{10}{\rm M_{\odot}}$ (417 halos) and the dash-triple-dotted
  line is for $2.1 \times 10^{10}{\rm M_{\odot}}$ (82 halos). The solid
  lines show the predictions of semi-analytical model (kindly provided
  by Claudio Firmani) for $10^{13}$, $10^{12}$, and $10^{11} {\rm
    M_{\odot}}$, from the bottom to the top.}
\label{fig:3}
\end{figure}

Unfortunately, there is no simple and straightforward way to assign a
mass for all halos identified in the simulation. Unlike the isolated
halos identified by HFOF at virial overdensity, the satellite halos,
although surviving, are subject to the tidal stripping which reduces
their mass. They are limited therefore by tidal, rather then virial,
radius. To assign masses to the halos we proceed as follows. The
isolated halos are assigned the mass inside the virial radius or radius
of $100 h^{-1}$ kpc, whichever is smaller.  The satellite halos are
assigned the total mass of {\em gravitationally bound} particles within
their tidal radius (or, again, within $100 h^{-1}$ kpc, whichever is
smaller). The tidal radius is determined as the radius at which the
density profile of a halo flattens (stops decreasing).

We now construct the complete mass evolution histories for the set of
{\em all} halos with the masses assigned as described above.  We have
divided these halos into five groups with masses $M_0 > 10^{13}{\rm
  M_{\odot}}$, $ 10^{13}{\rm M_{\odot}} > M_0 > 5 \times 10^{12}{\rm
  M_{\odot}}$, $ 5 \times 10^{12}{\rm M_{\odot}} > M_0 > 10^{12}{\rm
  M_{\odot}}$, $ 10^{12}{\rm M_{\odot}} > M_0 > 5 \times 10^{11}{\rm
  M_{\odot}}$, and $ 5 \times 10^{11}{\rm M_{\odot}} > M_0$. We defined
a subset of 3674 halos, mass of which increases (with allowance for
small statistical fluctuations) at all epochs.  As before, the mass of
these objects is normalized to their final mass at $z=0$.  The mass
evolution of these halos is shown in Fig. 4 (solid lines). The overall
evolution is similar to the mass evolution of isolated halos described
above (Fig. 3). Note, however, that while the mass evolution tracks are
curved in Fig.3, the mass evolution of the sample that includes
satellites can be better represented by the straight lines in these
log-log plots. This difference is due to the different halo selection
procedure and to the different assignment of mass to the selected
halos.

In the two lowest mass ranges we also find an additional subset of 2650 halos,
whose masses decrease after $z=1$ (dashed and dot-dashed curves in
Fig.4). Their mass increases at high redshifts, reaches a
clear maximum and decreases thereafter. The mass of these halos grows 
first due to accretion of surrounding material and of smaller halos.
At some point, however, these halos are being accreted by more
massive halos and they start to loose mass due to the tidal stripping
and interaction with other satellite halos. At $z=1$ these halos are
distributed similarly to the the rest. At $z=0$, however, they are
clustered more strongly than the overall halo population.

\begin{figure}
\centerline{\psfig{figure=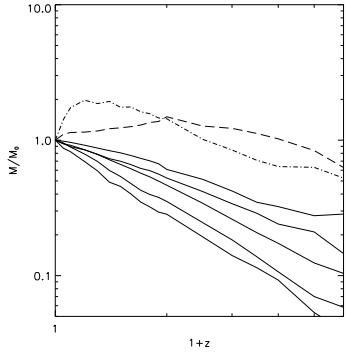,height=5cm}}
\caption[]{Mass evolution of halos. The solid lines are
  for average masses of (from the bottom to the top) $1.2 \times
  10^{13}{\rm M_{\odot}}$ (14 halos), $6.6 \times 10^{12}{\rm
    M_{\odot}}$ (34 halos), $1.9 \times 10^{12}{\rm M_{\odot}}$ (442
  halos), $7.0 \times 10^{11}{\rm M_{\odot}}$ (534 halos), and $2.4
  \times 10^{11}{\rm M_{\odot}}$ (2650 halos).  The
  dot-dashed  (average mass of $6.9 \times 10^{11}{\rm
    M_{\odot}}$) and the dashed (average mass of $2.0 \times
  10^{11}{\rm M_{\odot}}$) lines show the mass evolution of a subset of
  halos which loose mass between $z=1$ and the present due to the tidal
  stripping in groups and clusters.}
\label{fig:4}
\end{figure}

This is illustrated in Fig. 5, which shows the correlation function for
the two subsets of halos: always increasing mass and decreasing mass at
$z<1$.  The correlation functions of the former has a lower amplitude
and is not as steep as the correlation function of the latter. 
Note that the CF of the halos with the ever-increasing mass 
is anti-biased at scales $\lesssim 10 h^{-1}$ Mpc,
the CF of the halos that loose mass is actually positively biased. This
reflects the fact that the loosing mass halos are found within massive
systems such as massive galaxies, groups, and clusters, and are
therefore strongly clustered. 

One might speculate that this difference in the correlation functions may
serve as a possible explanation for the color segregation of the
correlation amplitude that has been recently observed (Carlberg et al.
1998). In fact, one could expect that the galaxies hosting halos which
undergo different mass evolution also show different properties, and
colors in particular. Further studies are necessary to test whether
this simple picture can really explain the observations. 

\begin{figure}
\centerline{\psfig{figure=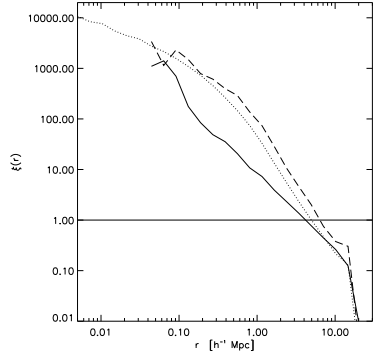,height=5cm}}
\caption[]{The correlation function of halos with circular velocities
$v_{circ} > 120$ km/s. The solid line corresponds to halos the mass of
which always increases (solid lines in Fig. 4), the dashed line
corresponds to halos which loose mass during evolution (dashed and
dashed-dotted lines in Fig. 4).  The correlation function of dark
matter particles is shown by the dotted line.  }
\label{fig:5}
\end{figure}

This work was funded by the NSF and NASA grants to NMSU.  SG acknowledges
support from Deutsche Akademie der Naturforscher Leopoldina with means
of the Bundesministerium f\"ur Bildung und Forschung grant LPD 1996. We
thank Claudio Firmani and Vladimir Avila-Reese for providing mass
evolution predictions of semi-analytical models. The numerical
simulations has been carried out at the Origin2000 computers at NCSA
and Naval Research Laboratory (NRL).

\end{document}